\documentclass[aps,prl,superscriptaddress,twocolumn,floatfix,showpacs]{revtex4-2}
\usepackage[utf8]{inputenc}
\usepackage[T1]{fontenc}
\usepackage{graphicx,color}
\usepackage[utf8]{inputenc}
\usepackage{amsmath,amssymb,bm,amsfonts,dsfont,mathrsfs,amsthm}
\usepackage{braket}
\usepackage{txfonts,comment}
\usepackage[colorlinks=true,linkcolor=blue,citecolor=blue,urlcolor=blue]{hyperref}
\usepackage{soul}
\usepackage{tikz}
\usetikzlibrary{shapes, arrows, positioning, calc, backgrounds}
\newtheorem{theorem}{Theorem}[]
\newcommand{\beginsupplement}
    {
    \setcounter{equation}{0}
    \setcounter{figure}{0}
    \setcounter{table}{0}
    \setcounter{section}{0}
    \makeatletter
    \renewcommand{\theequation}{S\arabic{equation}}
    \renewcommand{\thefigure}{S\arabic{figure}}
    \renewcommand{\bibnumfmt}[1]{[S##1]}
    }

\hypersetup{colorlinks=true, linkcolor=blue, citecolor=blue, urlcolor=blue}

\begin{document}

\author{Chiranjib Mukhopadhyay}
\affiliation{Institute of Fundamental and Frontier Sciences, University of Electronic Sciences and Technology of China, Chengdu 611731, China}
\affiliation{Key Laboratory of Quantum Physics and Photonic Quantum Information, Ministry of Education, University of Electronic Science and Technology of China, Chengdu 611731, China}
\author{Abolfazl Bayat}
\affiliation{Institute of Fundamental and Frontier Sciences, University of Electronic Sciences and Technology of China, Chengdu 611731, China}
\affiliation{Key Laboratory of Quantum Physics and Photonic Quantum Information, Ministry of Education, University of Electronic Science and Technology of China, Chengdu 611731, China}
\affiliation{Shimmer Center, Tianfu Jiangxi Laboratory, Chengdu 641419, China}

\author{Victor Montenegro}
\affiliation{Department of Physics, Khalifa University, Abu Dhabi 127788, United Arab Emirates}
\affiliation{Institute of Fundamental and Frontier Sciences, University of Electronic Sciences and Technology of China, Chengdu 611731, China}
\affiliation{Key Laboratory of Quantum Physics and Photonic Quantum Information, Ministry of Education, University of Electronic Science and Technology of China, Chengdu 611731, China}

\author{Matteo G.A. Paris}
\affiliation{Dipartimento di Fisica, Università di Milano, I-20133 Milan, Italy}

\title{Beating joint quantum estimation limits with stepwise multiparameter metrology}


\begin{abstract}
Conventional multiparameter quantum sensing relies on joint estimation, but this approach faces two key limitations: theoretical bounds may be unattainable due to measurement incompatibility, and sensing may fail due to parameter interdependencies. We propose stepwise estimation and identify regimes where it outperforms joint estimation. For multiple quantum sensors, this scheme achieves far lower error bounds than joint estimation. With many-body probes, stepwise sensing retains a quantum-enhanced scaling advantage often lost in joint estimation due to parameter correlations. We demonstrate its concrete advantages through Bayesian implementations across diverse examples.
\end{abstract}
\maketitle

\emph{Introduction.--}
Quantum systems, with their inherent sensitivity to external perturbations, are ideal for estimating field strengths~\cite{giovannetti2006quantum,giovannetti2011advances,xiang2013quantum, taylor2016quantum,toth2014quantum, degen2017quantum,polino2020photonic, demille2024quantum,mishra2021driving,montenegro2021global,mishra2022integrable,marciniak2022optimal,he2023stark,mukhopadhyay2025current,mukhopadhyay2024saturable,qvarfort2018gravimetry,montenegro2025heisenberg,kwon2020magnetic} and Hamiltonian parameters~\cite{campos2007quantum,zanardi2008quantum,shabani2011estimation,dooley2021robust,sarkar2022free,montenegro2022probing,mukhopadhyay2024modular,sahoo2024localization,ye2024essay,zhang2025quantum,burgarth2017evolution} with unparalleled precision. While the ultimate precision of quantum probes is well-established for single-parameter estimation, achieving comparable precision in multiparameter estimation with a single sensor remains a major challenge~\cite{demkowicz2020multi,szczykulska2016multi,liu2020quantum,meyer2021fisher}. The standard setting of multiparameter quantum estimation proceeds via joint estimation (JE) of parameters. The ultimate precision bound for JE is furnished by the standard Cram\'{e}r-Rao inequality lower bounding the total estimation error with the inverse of a certain family of matrices -- known as Fisher information matrices~\cite{petz2011introduction}. Two key problems emerge: (i) any near-singularity of these matrices rapidly degrade the sensing precision~\cite{mihailescu2025uncertain,goldberg2021taming,razavian2020quantumness,frigerio2024overcoming,yang2025overcoming,mihailescu2025metrological,he2025scrambling,mihailescu2024multiparameter,candeloro2024dimension}, moreover such singularities are not merely pathological but naturally arise in practical sensing situations with both many-body \cite{mihailescu2024multiparameter} and optical~\cite{peng2023spatial} probes as well as private quantum sensor networks~\cite{hassani2025privacy}; and (ii) in general, the bounds may not even be saturable in multiparameter sensing due to the incompatibility of optimal measurements for estimating each parameter individually~\cite{ragy2016compatibility,carollo2019quantumness,belliardo2021incompatibility,xia2023toward,albarelli2022probe,chen2022incompatibility}. Thus a natural question arises -- is there an alternative framework for multiparameter quantum metrology which evades the problem of incompatible measurements while also leading to significantly increased precision in the vicinity of the singularity? 



In this letter, we indeed propose a fundamentally distinct approach, termed stepwise estimation (SE), which decomposes the problem into a sequence of single-parameter estimations. We demonstrate that this SE scheme can yield a strictly lower total estimation error than conventional joint estimation (JE) strategies. Crucially, we validate these analytical advantages with concrete Bayesian implementations across a range of quantum sensing scenarios.


\emph{Multiparameter joint estimation scheme.--} Let us consider a probe state $\rho(\pmb{\lambda})$ encoding information about a family of unknown parameters $\pmb{\lambda}{=}\lbrace\lambda_i\rbrace$. In conventional joint estimation (JE) strategies \cite{szczykulska2016multi,liu2020quantum,gessner2018sensitivity,proctor2018multiparameter,nichols2018multiparameter,gessner2020multiparameter,goldberg2021intrinsic,albarelli2020perspective,meyer2021variational,rubio2020bayesian,di2022multiparameter}, we assume there is a single setting in which a quantum  measurement is performed over multiple rounds on the probe to simultaneously extract information about all the parameters. For concreteness, let us assume this measurement setting is given by positive operator-valued measure (POVM) elements $\pmb{\Pi} = \lbrace \Pi_\mu\rbrace$ with outcomes $\lbrace x_{\mu}\rbrace$ obtained with probabilities $p(x_{\mu}|\pmb{\lambda})= \text{Tr}[\Pi_{\mu} \rho(\pmb{\lambda})]$, respectively. The family of point estimators $\hat{\pmb{\Lambda}}$ corresponding to the parameters $\pmb{\lambda}$ are then constructed through post-processing of the measurement outcomes, which yield estimates $\pmb{\lambda}^{\text{est}}$, respectively. In this JE strategy performed over a total of $M$ rounds, the \textit{multiparameter Cram\'{e}r-Rao bound (MCRB)}~\cite{cramer1946contribution, rao1992information} furnishes a fundamental bound on the precision of estimation. Specifically, for the covariance matrix of unbiased estimators $\hat{\pmb{\Lambda}}$ denoted by $\text{Cov}[\hat{\pmb{\Lambda}}]_{ij}{=} \text{Tr}[\rho \hat{\Lambda}_i\hat{\Lambda}_j] - \text{Tr}[\rho \hat{\Lambda}_i]\text{Tr}[\rho \hat{\Lambda}_j]$, the MCRB takes the form of the matrix inequality
\begin{equation}
    \text{Cov}[\hat{\pmb{\Lambda}}] \ge \frac{\textbf{F}^{-1}(\pmb{\lambda})}{M} \ge \frac{\textbf{Q}^{-1}(\pmb{\lambda})}{M},
    \label{eq:mcrb}
\end{equation}
\noindent where $\textbf{F}(\pmb{\lambda})$ is the \textit{Fisher information matrix (FIM)} with elements $F_{ij}{=}\sum_{\mu} p(x_\mu|\pmb{\lambda}) \left(\partial_{\lambda_i} \ln p(x_\mu|\pmb{\lambda})  \right) \left(\partial_{\lambda_j} \ln p(x_\mu|\pmb{\lambda})  \right)$ and $\textbf{Q}(\pmb{\lambda})$ is the \textit{quantum Fisher information matrix (QFIM)} independent of individual measurement settings $\pmb{\Pi}$. For pure states $|\psi(\pmb{\lambda})\rangle$, the elements of $\mathbf{Q}$ can be obtained as $Q_{ij} {=} 4\ \text{Re}(\langle\partial_{\lambda_i}\psi|\partial_{\lambda_j}\psi\rangle {-} \langle\partial_{\lambda_i}\psi|\psi\rangle  \langle\psi|\partial_{\lambda_j}\psi\rangle)$. In Eq.\eqref{eq:mcrb}, the first inequality is saturable. However, the second inequality given by $\mathbf{Q^{-1}}$ is generally not tight unlike the case for single parameter estimation. Since matrix inequalities are often difficult to interpret in a straightforward way, one usually transforms the above version of MCRB into the following family of scalar inequalities by introducing a positive semidefinite weight matrix $\textbf{W}$ as, 
\begin{equation}
    \text{Tr}\left[\textbf{W}\text{Cov}(\hat{\pmb{\Lambda}})\right] \geq \frac{\text{Tr}\left[\textbf{W}\textbf{F}^{-1}(\pmb{\lambda})\right]}{M} \geq \frac{\text{Tr}\left[\textbf{W}\textbf{Q}^{-1}(\pmb{\lambda})\right]}{M}.
    \label{eq:mcrb_scalar}\nonumber
\end{equation}
In the rest of the paper, for simplicity we restrict ourselves to the case of two unknown parameters, i.e., $\pmb{\lambda} {=} \lbrace\lambda_1{,} \lambda_2\rbrace$. By setting $\textbf{W}$ to identity, namely $\textbf{W}{=}\mathbb{I}_2$, one assigns equal importance to both of the parameters. This results in
the following lower bound to sum of variances of the corresponding estimators $\hat{\pmb{\Lambda}} = \lbrace\hat{\Lambda}_1, \hat{\Lambda}_2\rbrace$
\begin{equation}
    \Delta_{\hat{\Lambda}_1}^2 + \Delta_{\hat{\Lambda}_2}^2 \ge \frac{\text{Tr} [\textbf{Q}^{-1}(\lambda_1,\lambda_2)]}{M} = \frac{\mu}{M},
    \label{eq:joint_strategy}
\end{equation}



While the above inequality may not be saturable in general, a saturable version of the MCRB known as Holevo Cram\'{e}r-Rao bound has been derived~\cite{holevo2006noncommutative}, and recently cast in the semidefinite programming formulation~\cite{albarelli2019evaluating}. However, explicit expressions for this lower bound are rare ~\cite{suzuki2016explicit,bressanini2024multi,chang2025multiparameter}. Note that the inequality in Eq.~\eqref{eq:joint_strategy} can be rewritten in the Holevo Cram\'{e}r-Rao form by replacing $\mu/M$ with $\mu_H/M$.\\

\emph{Multiparameter stepwise estimation scheme.--} In the stepwise estimation scheme, we divide the total $M$ rounds of measurement into two steps: (i) first $m_1$ rounds of measurements are used to construct the unbiased estimator $\hat{\Lambda}_1$ for estimating the parameter $\lambda_1$ which yields the value  ${\lambda}_1^{\text{est}}$; and (ii) armed with an estimate ${\lambda}_1^{\text{est}}$ for parameter $\lambda_1$, the rest  $m_2{=}M{-}m_1$ measurement rounds are devoted to the single parameter estimation problem of estimating the unknown parameter $\lambda_2$. In this setting, the precision bound for sum of variances of unbiased estimators $\hat{\Lambda}_1$ and $\hat{\Lambda}_2$ is set by the expression 
\begin{equation}
    \Delta_{\hat{\Lambda}_1}^2 + \Delta_{\hat{\Lambda}_2}^2 \geq \frac{\text{Tr} [\textbf{W}_1 \textbf{Q}^{-1} (\lambda_1,\lambda_2)]}{m_1} + \frac{1}{m_2 Q_{22} (\lambda_1^{\text{est}},\lambda_2)} = \frac{\mu'}{M},
    \label{eq:12_strategy}
\end{equation}
where the weight matrix $\textbf{W}_1{=}\text{diag}(1,0)$, and the single parameter QFI element $Q_{22}$ for the second parameter is predicated on the first parameter $\lambda_1$ being set to its estimate $\lambda_1^{\text{est}}$ obtained in the first step. Likewise, one can estimate $\lambda_2$ first spending $m_1$ rounds, followed by single-parameter estimation of $\lambda_1$ over next $m_2{=}M{-}m_1$ rounds. The precision bound for sum of variances of unbiased estimators $\hat{\Lambda}_1$ and $\hat{\Lambda}_2$ is then set by
\begin{equation}
    \Delta_{\hat{\Lambda}_1}^2 + \Delta_{\hat{\Lambda}_2}^2 \ge \frac{\text{Tr} [\textbf{W}_2 \textbf{Q}^{-1} (\lambda_1,\lambda_2)]}{m_1} + \frac{1}{m_2 Q_{11}(\lambda_1,\lambda_2^{\text{est}})} = \frac{\mu''}{M},
    \label{eq:21_strategy}
\end{equation}
with the weight matrix $\textbf{W}_1{=}\text{diag}(0,1)$, and the single parameter QFI element $Q_{11}$ for the parameter $\lambda_1$ being predicated on the second parameter $\lambda_2$ being set to its estimate $\lambda_2^{\text{est}}$ obtained in the first step. Thus, parameterizing $m_1 {=} M\gamma$ with $\gamma {\in} 
[0,1]$, the lowest precision bound among all possible SE schemes is given by $\tilde{\mu}/M$, where
\begin{equation}
 \tilde{\mu} {=} \min_{\gamma} (\mu'{,} \mu''), \hspace{0.5cm} \text{with} \hspace{0.5cm} \gamma_{\text{opt}}{=} \arg\min_{\gamma} (\mu'{,} \mu'')
\end{equation}
being the optimal measurement budgeting strategy. Assuming unbiased estimators, the corresponding estimates $\pmb{\lambda}^{\text{est}}$ all converge to their respective true values $\pmb{\lambda}$. Thus one can reasonably replace $Q_{11}(\lambda_1^{\text{est}}{,} \lambda_2)$ with $Q_{11}(\lambda_1{,}\lambda_2)$, and $Q_{22}(\lambda_1{,} \lambda_2^{\text{est}})$ with $Q_{22}(\lambda_1{,}\lambda_2)$ respectively in the asymptotic limit. In this asymptotic setting, the lower bound of estimation error of the SE scheme $\tilde{\mu}$ is guaranteed to lie below the lower bound of estimation error of the JE scheme $\mu$, i.e. $\tilde{\mu}{<}\mu$, when the following sufficiency condition is met
    \begin{equation}
       \frac{Q_{12}^2 (\lambda_1{,}\lambda_2)}{Q_{11} (\lambda_1{,}\lambda_2)Q_{22}(\lambda_1{,}\lambda_2)} > 2\sqrt{2} {-}2  \approx 0.828.
    \label{eq:two-parameter_condition}
    \end{equation}  
This is the first main result of this letter and has been proved in the supplemental material (SM). Note that, the criterion in inequality~\eqref{eq:two-parameter_condition} is a sufficiency condition, thus even if it is not satisfied, this does not necessarily imply $\tilde{\mu}{>}\mu$. Moreover, for a singular QFIM, the left hand side of Eq.~(\ref{eq:two-parameter_condition}) equals unity. Hence, satisfying the above inequality requires a QFIM sufficiently close to singular. In the SM, we have also provided a complementary necessary condition for the Holevo Cram\'er-Rao bound, i.e. $\tilde{\mu}{<}\mu_{H}$, for a class of so-called D-invariant probes~\cite{suzuki2018classification} where $\mu_H$ is explicitly computable. As an aside, this result also implies that $\tilde{\mu}{>}\mu$ for all QFIMs sufficiently close to diagonal. The inequality in Eq~\eqref{eq:two-parameter_condition} gives rise to two important questions. Firstly, what is the range of parameters $(\lambda_1,\lambda_2)$ over which the criterion of Eq.~(\ref{eq:two-parameter_condition}) is satisfied? Secondly, while satisfying this inequality shows that the SE bound is lower than the JE bound, can one truly beat the JE limit using a concrete estimation protocol such as Bayesian inference? In the following, we answer both these questions in the affirmative through specific examples.   

\begin{figure}
    \centering
    \includegraphics[width=\linewidth]{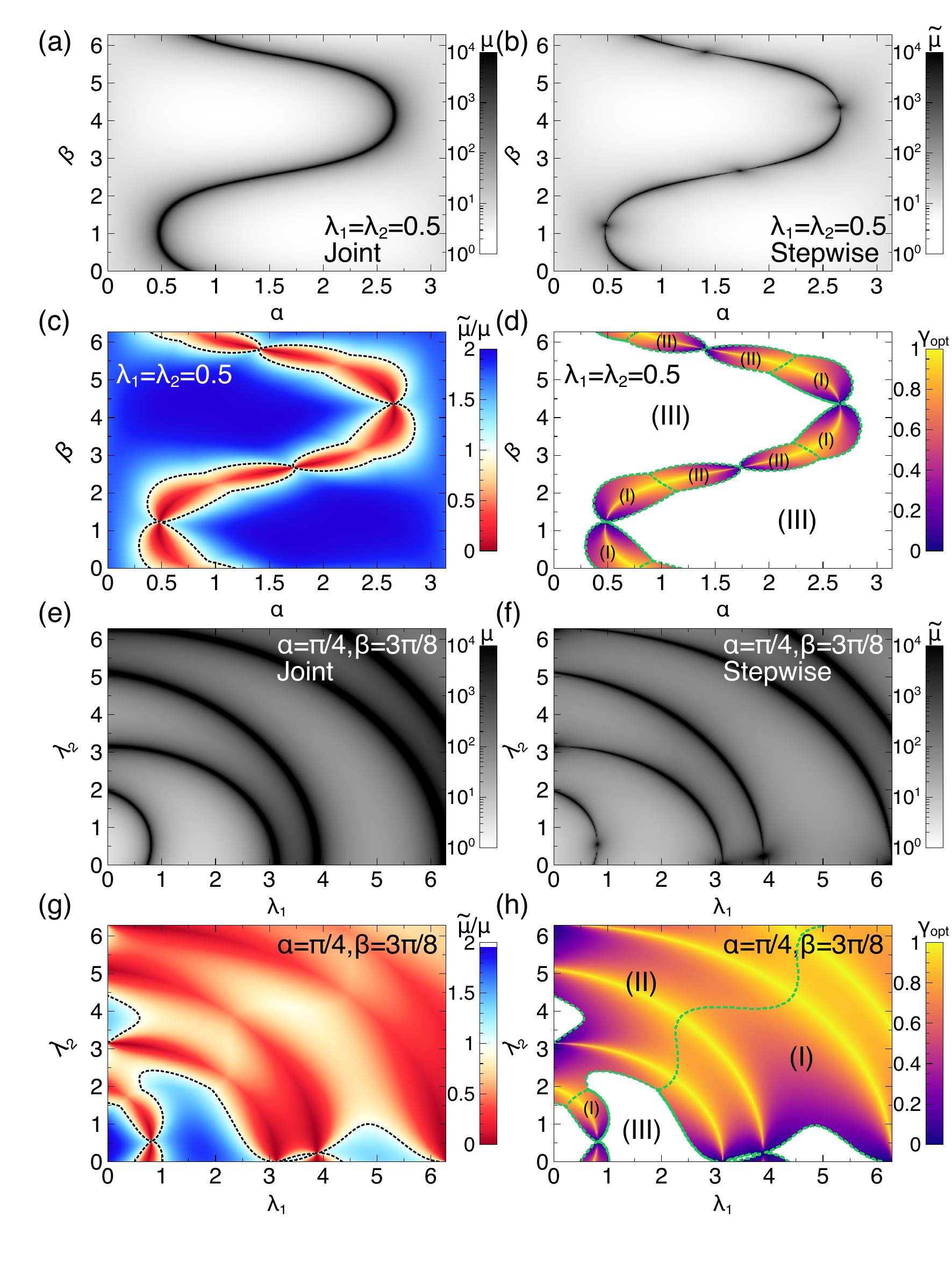}
    \caption{Qubit probe estimating $(\lambda_1, \lambda_2)$. (a) JE bound $\mu$ with initial conditions $(\alpha, \beta)$, where $(\lambda_1,\lambda_2){=}(0.5,0.5)$. (b) same for optimal SE bound $\tilde{\mu}$. (c) Ratio $\tilde{\mu}/\mu$, black dotted line separates regions where JE (blue) or SE (red) furnishes a better bound $(\tilde{\mu}/\mu {\gtrless} 1)$. (d) Optimal strategy $\gamma_{\text{opt}}$, green dotted lines separate regions (I) SE estimating $\lambda_1$ first, (II) SE estimating $\lambda_2$ first, (III) JE, furnishes a lower bound. (e)-(h) reproduces same figures as (a)-(d) with parameter values $(\lambda_1,\lambda_2)$, where initial condition $(\alpha,\beta){=}(\pi/4, 3\pi/8)$. }
    \label{fig:qubit}
\end{figure}

\emph{Qubit probe.--} As the first illustration, let us consider a qubit probe initially prepared in the state $|\psi_0\rangle {=} \cos \frac{\alpha}{2} |0\rangle {+} e^{i\beta}\sin \frac{\alpha}{2}|1\rangle$. The unknown parameters $\{ \lambda_1,\lambda_2\}$ are encoded in the quantum state through unitary operation $|\psi_{\text{probe}}\rangle{=}\hat{U}|\psi_0\rangle {=} e^{-i\left(\lambda_1\sigma^x + \lambda_2 \sigma^z \right)} |\psi_0\rangle$, with $\sigma^{x,z}$ being the Pauli operators. In Figs.\ref{fig:qubit}(a)-(b) we plot $\mu$ and $\tilde{\mu}$ respectively as a function of initial parameters $\alpha$ and $\beta$, keeping $\lambda_1{=}\lambda_2{=}0.5$, note that the QFIM is computed with respect to $\{\lambda_1,\lambda_2\}$. The dark region indicates where the QFIM is nearly singular, leading to significantly larger estimation errors. In Fig.\ref{fig:qubit}(c) we plot the corresponding ratio $\tilde{\mu}{/}\mu$ as a function of initial parameters $\alpha$ and $\beta$, keeping $\lambda_1{=}\lambda_2{=}0.5$. The red region, encircled by dotted lines, denotes where the SE bound is below the JE bound, i.e. $\tilde{\mu}{/}\mu{<}1$. In Fig.~\ref{fig:qubit}(d) we specify the optimal measurement budgeting parameter $\gamma{=}\gamma_{\rm opt}$ for the SE scenario. In addition, we identify three regions: (i) the region I, where the optimal strategy is SE measuring $\lambda_1$ first, (ii) the region II, where the optimal strategy is SE measuring $\lambda_2$ first, and (iii) where JE bound $\mu$ outperforms the optimal SE bound $\tilde{\mu}$. In Figs.\ref{fig:qubit}(e)-(f) we plot $\mu$ and $\tilde{\mu}$ respectively as a function of unknown parameters $\lambda_1$ and $\lambda_2$, keeping initial parameters fixed. Similar to panels (a) and (b), the dark regions indicate where QFIM is nearly singular. In Fig.~\ref{fig:qubit}(g) we plot the ratio $\tilde{\mu}{/}\mu$ as a function of unknown parameters, keeping $\alpha$ and $\beta$ fixed. Again the region inside the dotted lines is where the SE bound outperforms the JE bound. In Fig.~\ref{fig:qubit}(h), the optimal budgeting strategy $\gamma_{\rm opt}$ and the corresponding regions are indicated in $\{ \lambda_1,\lambda_2\}$ space. Thus, this simplest qubit probe example already demonstrates the potential of SE as a superior tool for multiparameter quantum sensing in the region where sensing is challenging due to nearly singular QFIM, according to Eq.~(\ref{eq:two-parameter_condition}).

\begin{figure}
    \centering
    \includegraphics[width=\linewidth]{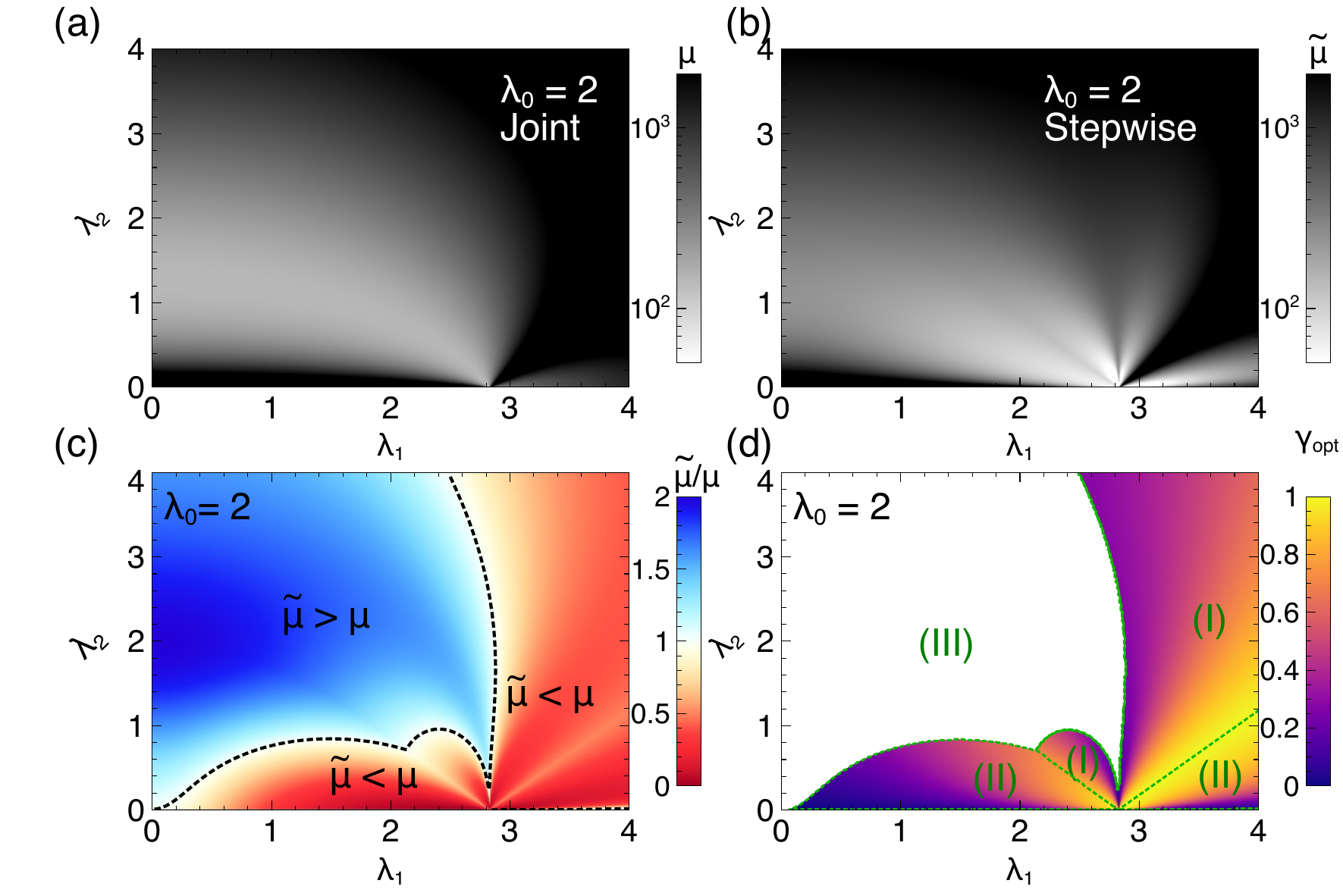}
    \caption{Three-level LZ probe estimating $(\lambda_1, \lambda_2)$. (a) JE bound $\mu$ with $(\lambda_1,\lambda_2)$, (b) same for optimal SE bound $\tilde{\mu}$. (c) Ratio $\tilde{\mu}/\mu$, black dotted line separates regions where JE (blue) or SE (red) furnishes a better bound $(\tilde{\mu}/\mu{\gtrless} 1)$. (d) Optimal strategy $\gamma_{\text{opt}}$, green dotted lines separate regions (I) SE estimating $\lambda_1$ first, (II) SE estimating $\lambda_2$ first, (III) JE, furnishes a lower bound. $\lambda_0{=2}$ throughout.}
    \label{fig:lz_plot}
\end{figure}
 
\emph{Three-level Landau-Zener probe.--} Beyond qubit probes, a natural choice for direct demonstration of quantum-enhanced sensitivity is many-body probes employing criticality associated with gap-closing. See Ref.~\cite{montenegro2024review} for a review. To proceed along this direction, let us first consider a minimal toy model multiparameter example exhibiting gap-closing~\cite{}, namely the three-level Landau-Zener (LZ) Hamiltonian given by \cite{carroll1986generalisation,shevchenko2010landau,band2019three} $H = \lambda_0 \left(|0\rangle\langle 0| {-} |2\rangle\langle2|\right) {+} \lambda_1\left(|0\rangle\langle1| {+} |1\rangle\langle0|\right) {+} \lambda_2\left(|1\rangle\langle2| {+}|2\rangle\langle1|\right)$, where $(\lambda_1{,} \lambda_2)$ are transition parameters to be estimated. 
Transitions between the first and third energy levels are forbidden due to selection rules, with the driving parameter $\lambda_0$ typically varying with time. However, to simplify matters, we consider the instantaneous ground state of this LZ Hamiltonian, where $\lambda_0$ is set to be $\lambda_0{=}2$. For this choice the gap closing takes place at $(\lambda_1,\lambda_2){=}(2\sqrt{2},0)$. For the sake of clarity, in Figs.~\ref{fig:lz_plot}(a)-(b) we plot $\mu$ and $\tilde{\mu}$ as function of $\{\lambda_1,\lambda_2\}$, respectively. To gain more information, in  Fig.~\ref{fig:lz_plot}(c) we depict the ratio $\tilde{\mu}{/}\mu$ in the plane of $\{\lambda_1,\lambda_2\}$. The red parts, denoted by dotted lines, are the regions over which the SE bound is lower than the JE bound, i.e. $\tilde{\mu}{/}\mu{<}1$. In Fig.~\ref{fig:lz_plot}(d), the optimal budgeting ratio $\gamma_{\rm opt}$ and the three corresponding regions are shown. As the figure shows, at the vicinity of the gap closing point, the optimal strategy changes abruptly with parameter shifts. As evidenced by this example, there also exist several distinct parameter regimes in which SE outperforms JE, provided that optimal resource budgeting and measurement strategies are employed. \\

\begin{figure}
    \centering
    \includegraphics[width=\linewidth]{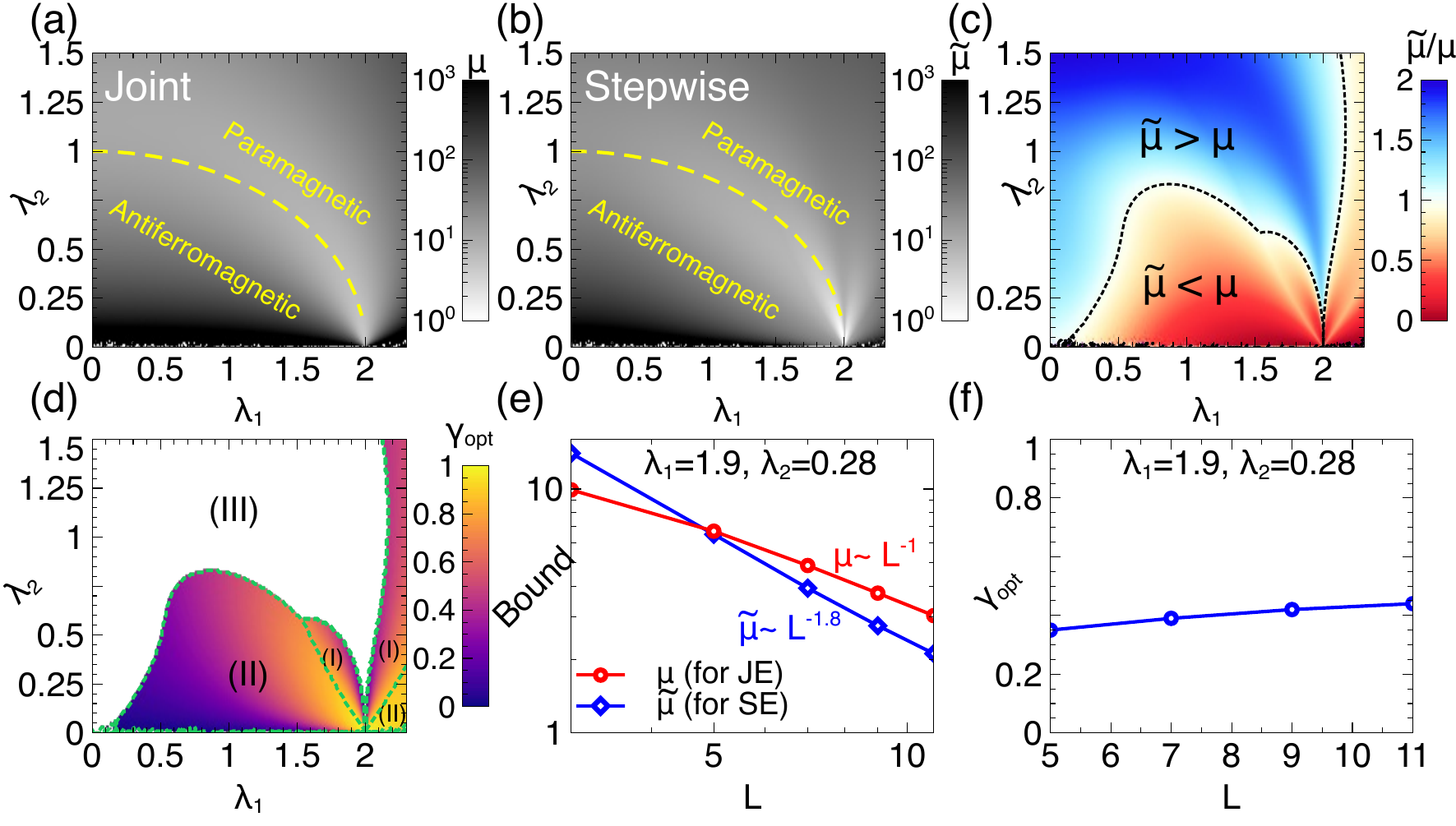}
    \caption{Mixed Ising probe estimating $(\lambda_1,\lambda_2)$. (a) JE bound $\mu$ with $(\lambda_1,\lambda_2)$, (b) same for optimal SE bound $\tilde{\mu}$. Yellow dashed lines are phase boundaries. (c) Ratio $\tilde{\mu}/\mu$, black dotted line separates regions where JE (blue) or SE (red) furnishes a better bound $(\tilde{\mu}/\mu \ {\gtrless} 1)$. (d) Optimal strategy $\gamma_{\text{opt}}$, green dotted lines separate regions (I) SE estimating $\lambda_1$ first, (II) SE estimating $\lambda_2$ first, (III) JE, furnishes a lower bound. (e) JE (red) and optimal SE (blue) bounds with size $L$ near first-order critical point at $(\lambda_1,\lambda_2) {=} (1.9,0.28)$, (f) Optimal strategy $\gamma_{\text{opt}}$ of estimating $\lambda_1$ first with size $L$ for $(\lambda_1,\lambda_2) {=} (1,9,0.28)$. $L{=}6$ for (a)-(d). Periodic boundary conditions assumed throughout. 
    }
    \label{fig:many-body}
\end{figure}

\emph{Many-body mixed Ising probe.--}  Let us now consider a genuine many-body system, viz., a mixed Ising probe determined by the Hamiltonian $H {=} J \sum_{i} \sigma_i^x \sigma_{i+1}^x  {-} \sum_{i} \left[h_{x}\sigma_{i}^x {+ }h_z \sigma_{i}^z \right]$, where $J$ is the exchange coupling and  $h_x$ ($h_z$) is the longitudinal (transverse) magnetic field. The probe is considered to operate at its ground state.  The goal is to estimate $\lambda_1{=}h_x{/}J$ and $\lambda_2{=}h_z{/}J$. In the absence of the longitudinal field $h_x$, the conventional transverse Ising model has a quantum phase transition at $\lambda_2 {=}{\pm} 1$ between ordered ($\lambda_2{<}1$) and disordered ($\lambda_2{>}1$) phases. In the absence of the transverse field $h_z$, the conventional Ising model has a first-order phase transition at $\lambda_1{=}2$. If both fields are active, a critical line interpolating between these limiting cases separates the phase boundaries in the $\{\lambda_1,\lambda_2\}$ plane. 
We investigate both JE and SE strategies for estimating $\{\lambda_1,\lambda_2\}$ in the ground state of the Hamiltonian. In Figs.~\ref{fig:many-body}(a)-(b) we plot $\mu$ and $\tilde{\mu}$ as a function of $\lambda_1$ and $\lambda_2$, respectively with the phase boundaries overlaid. To elaborate, in Fig.~\ref{fig:many-body}(c) we plot the ratio $\tilde{\mu}{/}\mu$ as a function of $\{\lambda_1,\lambda_2\}$. The red region indicates the area in which SE bound is lower than the JE bound ($\tilde{\mu}{/}\mu{<}1$). The measurement budget parameter $\gamma_{\rm opt}$ and the corresponding three regions are plotted in Fig.~\ref{fig:many-body}(d). Interestingly, one can identify two distinct behaviors around the critical points. Near the quantum critical point $(\lambda_1,\lambda_2){=}(0,1)$, SE does not seem to afford any advantage. However, at the vicinity of the other critical point $(\lambda_1,\lambda_2){=}(2,0)$, the optimal strategy cycles abruptly between two SE strategies and JE, which is exactly reminiscent of the previously considered LZ probe near its gap closing point. Even more remarkably, Fig.~\ref{fig:many-body}(e) shows that at a point $(\lambda_1,\lambda_2){=}(1.9,0.28)$ near the latter critical point, this advantage of SE vis-a-vis JE grows scalably with system size $L$, with the former decreasing as ${\sim}L^{-1.8}$ outweighing the shot noise scaling ${\sim}L^{-1}$ exhibited by the JE strategy near the first-order critical point. Fig.~\ref{fig:many-body}(f) shows that the optimal strategy approaches $\gamma_{\text{opt}} {\approx} 0.5$ for every length $L$. Notice that although the individual QFIM  elements exhibit superlinear scaling near criticality, near-singular QFIM of the model prevents JE (but not SE) from attaining quantum enhanced sensitivity. This result shows that while the criticality-driven quantum enhancement may be lost when the QFIM is near singular around the phase boundary, the quantum advantage can actually be restored through our SE protocol. This constitutes the second result of this letter. \\


\begin{figure}
    \centering
    \includegraphics[width=\linewidth]{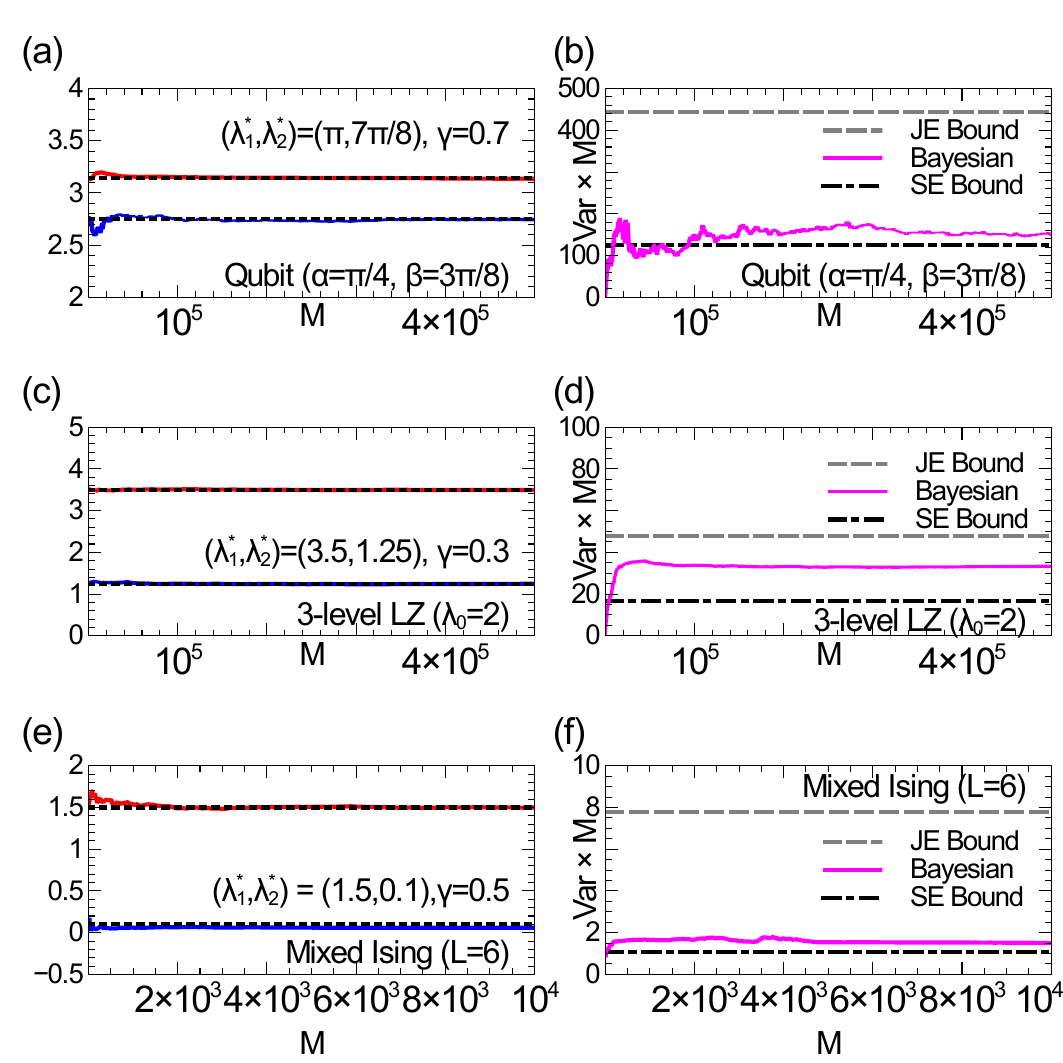}
    \caption{Bayesian SE for (a)-(b) qubit probe estimating $(\lambda_1,\lambda_2)$ with true values $(\pi,7\pi/8)$ where $(\alpha,\beta){=}(\pi/4,3\pi/8)$; (c)-(d) three-level LZ probe estimating $(\lambda_1,\lambda_2)$ with true values $(3.5,1.25)$ where $\lambda_0{=}2$; (e)-(f) mixed Ising probe estimating $(\lambda_1,\lambda_2)$ with true values $(1.5,0.1)$ where $L{=}6$, for total number of simulated measurements $M$ and budgeting $\gamma$. Left panels (a),(c),(e) depict convergence of Bayesian estimates given by red (blue) solid lines for $\lambda_1$($\lambda_2$) to true values (black dotted lines). Right panels (b),(d),(f) depict sum of Bayesian variances multiplied by $M$ (magenta solid line) vs. JE bound (gray dashed line) and SE bound (black dash-dotted line). In each case, uniform priors are chosen with width $\pi/5$ (qubit), $0.2$ (three-level LZ), $1$ (mixed Ising) respectively for both parameters. In (a)-(d), $\lambda_1$ is measured first and in (e)-(f), $\lambda_2$ is measured first.}
    \label{fig:bayesian}
\end{figure}

\emph{Bayesian Implementation.--}So far, we have shown that SE bound can outperform the JE bound, namely $\tilde{\mu}{<}\mu$, within a region of parameters which is model dependent.  However, since we are comparing bounds, one may at this point legitimately wonder whether this lower bound genuinely translates into better sensing precision for SE. To confirm this claim, we now illustrate the Bayesian implementation of SE for all the above examples. In this section we present the main results and provide the technical details in the SM. Note that through only relying on intuitive choices of fixed measurement bases, one can indeed demonstrate the superiority of SE strategy over the JE bound, which itself may not even be reachable. For this demonstration, we sample the true parameter values from the region where the SE bound indeed outperforms the JE bound (i.e., $\tilde{\mu}/\mu {<}1$), and follow a measurement budgeting strategy $\gamma$ close to the theoretically optimal one obtained via analyzing the bound $\tilde{\mu}$. For the qubit case, we employ measurement along $x$-basis, i.e, projectors $\lbrace |+\rangle, |-\rangle\rbrace$ for estimating $\lambda_1$ and measurement along $z$-basis, i.e, projectors $\lbrace |0\rangle, |1\rangle\rbrace$ for estimating $\lambda_2$.  The convergence of the two parameters in the SE setting is shown in Fig.~\ref{fig:bayesian}(a). To compare with the JE bound, in  Fig.~\ref{fig:bayesian}(b) we plot the sum of the variances multiplied by the total number of measurements $M$ as a function of $M$. As the figure shows we can almost reach the SE bound and significantly outperform the JE bound. 
For the  three-level LZ probe, we  choose a natural measurement setting $\mathbf{\Pi}_{1}{=}\lbrace (1,1,0)^T, (1,-1,0)^T, (0,0,1)^T\rbrace$ for estimating $\lambda_1$, and likewise $\mathbf{\Pi}_{2}{=}\lbrace (0, 1,1)^T, (0,1,-1)^T, (1,0,0)^T\rbrace$ for estimating $\lambda_2$ respectively. The convergence of the SE setting is shown in Fig.~\ref{fig:bayesian}(c). In Fig.~\ref{fig:bayesian}(d), we plot the sum of the variances multiplied by the total number of measurements $M$ as a function of $M$, which lies  between the SE and the JE bounds, exhibiting concrete advantage of  SE over  JE.  For the many-body mixed Ising model, we consider a probe of size $L{=}6$ and  measure total magnetization operators along $z$ and $x$-directions to estimate $\lambda_2$ and then $\lambda_1$, respectively. In  Fig.~\ref{fig:bayesian}(e) we plot the convergence of our estimation as a function of $M$. In Fig.~\ref{fig:bayesian}(f), we depict the sum of the variances multiplied by the total number of measurements $M$ as a function of $M$, which almost reaches the SE bound.  These examples all clearly show significant improvement achieved by SE through a concrete estimation protocol. 
This experimental-friendly Bayesian implementation of SE scheme confirming is the third main result of this letter. \\

\emph{Conclusions.--} In this letter, we propose stepwise estimation (SE) as a viable and fundamentally distinct approach to multiparameter quantum sensing. Our SE strategy outperforms conventional joint estimation (JE) schemes in several cases, and proves especially advantageous in situations where the QFIM becomes nearly singular.  We have three main results. Firstly, we analytically establish a sufficient criterion for the SE lower bound on achievable precision to lie significantly below the ultimate, yet generally unreachable, bound set by JE protocols. This assertion is backed through explicit treatment of multiple probes. Secondly, we identify many-body probes whose criticality-enhanced sensitivity is diminished for JE due to near-singularity of QFIM but is restored through the SE approach. Thirdly, we explicitly demonstrate that SE approach indeed results in lower estimation error through a concrete Bayesian implementation than the corresponding ultimate yet generally unreachable JE error bounds. Intuitively, the SE approach trades the error induced in JE due to quantum incompatibility for the classical error induced in imprecise estimation of the first parameter. We emphasize that nearly singular multi-parameter models may routinely arise in sensing with impurity probes \cite{mihailescu2024multiparameter}, sensing spatial modulation of a parameter \cite{peng2023spatial}, or in establishing a private network of sensors \cite{hassani2025privacy}, thus marking the SE scheme as a natural fit for quantum multiparameter metrology in these settings.

\emph{Acknowledgments.--}  We acknowledge support from the National Natural Science Foundation of China (Grants No. 12050410253, No. 92065115, No. 12274059, No. W2432005 and  No. 12374482) and from MUR - NextGenerationEU (Projects G53D23001110006-RISQUE, and G53D23006270001-QWEST).

\bibliography{stepwise}

\onecolumngrid    

\beginsupplement
\section{Supplemental Material}




\subsection{Brief Review of Multiparameter Quantum Estimation}

In the \textit{joint estimation} (JE) setting, the precision bound  for sum of variances of the two estimators $\hat{\Lambda}_1,\hat{\Lambda}_2$ of parameters $\lambda_1, \lambda_2$ respectively is set by the \textit{multiparameter Cram\'{e}r-Rao bound}~\cite{cramer1946contribution, rao1992information}
\begin{equation}
    \Delta_{\hat{\Lambda}_1}^2 + \Delta_{\hat{\Lambda}_2}^2 > \frac{\text{Tr} [\textbf{Q}^{-1}]}{M} = \frac{\mu}{M},
    \label{eq:joint_strategy_companion}
\end{equation} where $\textbf{Q}$ is the so-called \textit{quantum Fisher information matrix} (QFIM). The elements of this matrix are denoted by $\textbf{Q} {=} \lbrace Q_{ij} \rbrace$ with indices $i{,}j {=} 1{,}2$ corresponding to parameters $\lambda_1$ and $\lambda_2$ respectively. For a pure probe state $|\psi(\lambda_1,\lambda_2)\rangle$, QFIM elements are given by the expression 
\begin{equation}
    Q_{ij} = 4\ \text{Re}\left[\langle\partial_{\lambda_i}\psi|\partial_{\lambda_j}\psi\rangle - \langle\partial_{\lambda_i}\psi|\psi\rangle  \langle\psi|\partial_{\lambda_j}\psi\rangle\right]
\end{equation}

\noindent However, unlike the single-parameter estimation problem, this bound is generally not tight. Holevo found the following asymptotically tight bound $\mu_H$ for sum of variances of the two estimators $\hat{\Lambda}_1,\hat{\Lambda}_2$ of parameters $\lambda_1, \lambda_2$ respectively as \cite{holevo2006noncommutative}

\begin{equation}
    \Delta_{\hat{\Lambda}_1}^2 + \Delta_{\hat{\Lambda}_2}^2  \geq \mu_{H} = \min_{} h(\vec{\lambda}) [\hat{\textbf{X}}], 
\end{equation}
\noindent where $ \hat{\textbf{X}}{=}( \hat{X}_1, \hat{X}_2, . . . , \hat{X}_d)$ is a vector of Hermitian operators satisfying the locally unbiased conditions

\begin{align}
    \text{Tr} [\rho(\vec{\lambda}) \hat{X}_i] = 0 ~ \forall i \\
    \text{Tr} [\partial_{\lambda_i}\rho(\vec{\lambda}) \hat{X}_j] = \delta_{ij} ~ \forall i,j \\
\end{align}
\noindent and the function to minimize reads
\begin{equation}
  h(\vec{\lambda}) [\hat{\textbf{X}}] = \text{Tr}\left[\text{Re} \left( \textbf{Z}(\vec{\lambda}) [\hat{\textbf{X}}] \right) \right]  + \lVert \text{Im}\left( \textbf{Z}(\vec{\lambda}) [\hat{\textbf{X}}] \right) \rVert_{1}
\end{equation}
\noindent where $\lVert.\rVert_1$ is the usual matrix $l_1$-norm and elements of the matrix $\textbf{Z}(\vec{\lambda}) [\hat{\textbf{X}}]$ are given by 

\begin{equation}
    \left(\textbf{Z}(\vec{\lambda}) [\hat{\textbf{X}}]\right)_{ij} = \text{Tr}\left[\rho(\vec{\lambda}) \hat{X}_{i} \hat{X}_{j}\right] = \hat{\textbf{X}}^T S(\vec{\lambda})~\hat{\textbf{X}} 
\end{equation}

\noindent Albarelli \emph{et al} showed in Ref.~\cite{albarelli2019evaluating} that this functional minimization can be cast as a semi-definite program that can then be solved numerically. However, for certain simple systems, explicit analytical expressions are possible to be obtained. In particular, let us note that for $D$-invariant models \cite{suzuki2018classification}, the Holevo Cram\'{e}r Rao bound (HCRB) is written down as $\mu_{H} = \mu + \lVert \textbf{Q}^{-1} \textbf{D} \textbf{Q} \rVert_1$, where $\textbf{D}$ is the Uhlmann curvature matrix. The QFIM $\textbf{Q}$ and the Uhlmann curvature $\textbf{D}$ can in turn be interpreted as the real and imaginary components of the quantum geometric tensor $\textbf{G} = \textbf{Q} {+} i\textbf{D}$. For generic two-parameter qubit statistical models, an explicit formula for HCRB  was derived by Suzuki \cite{suzuki2016explicit} as 
\begin{eqnarray}
\mu_H = \begin{cases}
\mu_{R} & \text{if}~ \mu_{R} \geq \mu +\frac{ \lVert \textbf{Q}^{-1} \textbf{D}\textbf{Q} \rVert_1}{2} \\
\mu_{\text{R}} + \frac{\mu - \mu_{R} + \frac{1}{2}  \lVert \textbf{Q}^{-1} \textbf{D}\textbf{Q} \rVert_1 }{\mu - \mu_{R} +  \lVert \textbf{Q}^{-1} \textbf{D}\textbf{Q} \rVert_1} & \text{otherwise}
\end{cases}
\label{eq:generic_two_parameter_qubit_holevo}
\end{eqnarray}

Clearly, the RLD bound $\mu_{R}$ plays a crucial role in the achievable multiparameter estimation strategy with joint estimation. Let us further note that for pure state two-parameter models, the Uhlmann curvature matrix is defined as 
\begin{equation}
    \textbf{D} = 4 \begin{bmatrix} 0 & \text{Im} \left( \langle \partial_{\lambda_1} \psi | \partial_{\lambda_2}\psi \rangle \right) \\
     - \text{Im} \left( \langle \partial_{\lambda_1} \psi | \partial_{\lambda_2}\psi \rangle \right) & 0
    \end{bmatrix} = i \delta \sigma_y, 
    \label{eq:uhlmann}
\end{equation}
\noindent where $\delta = 4~ \text{Im} (\langle \partial_{\lambda_1} \psi| \partial_{\lambda_2}\psi\rangle)$, and $\sigma_y$ is one of the well known Pauli spin matrices.

\subsection{General two-parameter stepwise quantum estimation}

Armed with the review of joint measurement strategies for multiparameter estimation, let us now commence on the original contribution of this paper. Firstly, let us note the key source of difficulty with multiparameter estimation compared to the single parameter case where the QCRB, at least in the asymptotic limit, is a saturable bound. On a fundamental level, this difficulty arises from the incompatibility of optimal measurement bases for the many parameters to be estimated. The single optimal joint measurement basis for multiparameter estimation thus necessarily has to be a compromise between these strategies. In contrast, we concentrate on our \emph{stepwise estimation} (SE) scheme, where we break down the $M$ rounds of measurement in Eq.~\eqref{eq:joint_strategy_companion} into two sub-rounds $m_1{=}M\gamma$ and $m_2{=}M(1{-}\gamma)$ sequentially. For the first $m_1$ rounds, we focus on finding a good estimate of one parameter, say $\lambda_1$. Then armed with the estimate of this parameter, we spend the next $m_2$ rounds to optimize for estimation of the other parameter $\lambda_2$. Thus effectively we turn the problem of multiparameter estimation into a sequence of single parameter estimation problems.  We also keep in mind that this is a local estimation problem, the rough values of $\vec{\lambda} = \lbrace \lambda_1, \lambda_2 \rbrace$ are already available to us.  Let us start by denoting the two-parameter QFIM as
\begin{equation}
    \textbf{Q} = \begin{pmatrix}
        Q_{11} & Q_{12} \\
        Q_{12} & Q_{22} 
    \end{pmatrix}
\end{equation}

\noindent Then $\tilde{\textbf{Q}} = \textbf{Q}^{-1}$ is given by 

\begin{equation}
    \tilde{\textbf{Q}} = \frac{1}{Q_{11} Q_{22} - Q_{12}^2}\begin{pmatrix}
        Q_{22} & -Q_{12}\\
        -Q_{12} & Q_{11}
    \end{pmatrix}
\end{equation}

\noindent Now assume $M$ rounds of joint estimation is compared to the situation where we spend $M\gamma$ rounds of estimating observable 1, and then armed with its value, we spend next $M-M\gamma$ rounds of observable 2, where $\gamma \in [0,1]$. For joint estimation after $M$ rounds, the uncertainty is lower bounded by 
\begin{equation}
    \Delta_{\hat{\Lambda}_1}^2 + \Delta_{\hat{\Lambda}_2} ^2 > \frac{1}{M} \text{Tr}[\tilde{\textbf{Q}}] = \frac{1}{M} \frac{Q_{11} + Q_{22}}{Q_{11}Q_{22}-Q_{12}^2}
    \label{eq:joint}
\end{equation}

\noindent For stepwise estimation, first let us assume that first $M\gamma$ rounds are spent trying to estimate the first parameter and then $M-M\gamma$ rounds are spent trying to estimate the second parameter. In this case, the uncertainty is lower bounded by 
\begin{equation}
    \Delta_{\hat{\Lambda}_1}^2 + \Delta_{\hat{\Lambda}_2} ^2 > \frac{\text{Tr}[\textbf{W}_1 \textbf{Q}^{-1}]}{M\gamma} + \frac{1}{M-M\gamma}\frac{1}{Q_{22}},
    \label{eq:first1then2}
\end{equation}

\noindent where the weight matrix $\textbf{W}_1 {=}\text{diag}(1,0)$. Now, let us reverse the strategies, i.e., trying to estimate the second parameter first and then armed with its estimator trying to esimate the first parameter. In this case, denoting the weight matrix $\textbf{W}_2 {=}\text{diag}(0,1)$, the uncertainty is lower bounded by 

\begin{equation}
    \Delta_{\hat{\Lambda}_1}^2 + \Delta_{\hat{\Lambda}_2} ^2 > \frac{\text{Tr}[\textbf{W}_2 \textbf{Q}^{-1}]}{M\gamma} + \frac{1}{M-M\gamma}\frac{1}{Q_{11}}
    \label{eq:first2then1}
\end{equation}

\subsection{Sufficiency Condition for stepwise strategy to be better than joint Strategy}

\begin{theorem}
    For any two-parameter estimation problem, SE is guaranteed to outperform JE when the following condition on elements of QFIM \textbf{Q} is met 
    \begin{equation}
        \left(\frac{Q_{12}}{Q_{11}}\right)\left(\frac{Q_{12}}{Q_{22}}\right) \geq 2\sqrt{2} - 2 \approx 0.828 
    \label{eq:sufficiency_se_better}
    \end{equation}
\end{theorem}

\emph{Proof.--} If joint measurement is better than \emph{any} such stepwise strategy, then RHS of Eq.~\eqref{eq:joint} must be smaller than RHS of Eq.~\eqref{eq:first1then2} and Eq.~\eqref{eq:first2then1} for all $\gamma \in [0,1]$. For otherwise that strategy will give a bound lower than the joint estimation strategy. That is, the following two conditions have to be simultaneously met for all $\gamma \in [0,1]$

\begin{eqnarray}
    \frac{Q_{11} + Q_{22}}{Q_{11}Q_{22}-Q_{12}^2} \leq  \left[ \frac{Q_{22}/\gamma}{Q_{11}Q_{22}-Q_{12}^2}+ \frac{1}{(1-\gamma) Q_{22}} \right] \\
    \frac{Q_{11} + Q_{22}}{Q_{11}Q_{22}-Q_{12}^2} \leq \left[ \frac{Q_{11}/\gamma}{Q_{11}Q_{22}-Q_{12}^2}+ \frac{1}{(1-\gamma) Q_{11}} \right]
\end{eqnarray}

\noindent These conditions can be alternately written as 
\begin{eqnarray}
    \frac{Q_{11} - (\frac{1}{\gamma}-1)Q_{22}}{Q_{11}Q_{22} - Q_{12}^2} \leq \frac{1}{(1-\gamma)Q_{22}} \\
     \frac{-(\frac{1}{\gamma}-1)Q_{11} + Q_{22}}{Q_{11}Q_{22} - Q_{12}^2} \leq \frac{1}{(1-\gamma)Q_{11}}
\end{eqnarray}

\noindent Summing them together gives 

\begin{equation}
   \left(2-\frac{1}{\gamma}\right) \frac{Q_{11} +Q_{22}}{Q_{11}Q_{22} - Q_{12}^2} \leq \frac{1}{1-\gamma} \frac{Q_{11}+ Q_{22}}{Q_{11}Q_{22}}
   \label{eq:summed_up}
\end{equation}

\noindent After some algebra this leads to \begin{eqnarray}
    \frac{Q_{11} Q_{22} - Q_{12}^2}{Q_{11}Q_{22}} \geq 3- \left(2\gamma + \frac{1}{\gamma} \right)
\end{eqnarray}

\noindent Now, $2\gamma {+} 1/\gamma$ is bounded below by $2\sqrt{2}$. Thus, in any problem where $\frac{Q_{11} Q_{22} {-}Q_{12}^2}{Q_{11}Q_{22}} \leq 3{-} 2\sqrt{2}$ is where stepwise estimation will have a better result than joint estimation. That is, stepwise estimation is always guaranteed to succeed when 

\begin{equation}
    \left(\frac{Q_{12}}{Q_{11}}\right)\left(\frac{Q_{12}}{Q_{22}}\right) \geq 2\sqrt{2} - 2 \approx 0.828   
\end{equation} 
This completes the proof of the theorem. \qed \\

\noindent Note that this is a sufficiency condition, not a necessary condition for the stepwise estimation to be better since this ultimate JE estimation bound may indeed not be attainable. To have a more concrete picture of the performance of the stepwise strategy vis-a-vis joint estimation, one should consider the asymptotically achievable Holevo version of the CRB for the latter.

\subsection{When is joint measurement better than stepwise ?}

In the previous theorem, we presented the sufficiency condition for SE to better than JE. This effectively turns out to be the case for QFI matrix to be close to singular. What happens in the opposite limit ? We shall now consider the general case where the QFI matrix is close to diagonal and show that in this limit, achievable version of JE schemes outperform SE for a large class of quantum statistical models. That is, the following theorem holds. 

\begin{theorem}
    If the QFI matrix is close to diagonal, then for $D$-invariant models, the achievable version of JE, i.e., the Holevo version of CRB always outperforms  any stepwise strategy.
\end{theorem}

\emph{Proof.--} If no stepwise strategy can beat the achievable JE strategy via the HCRB, then there exists no $\gamma \in [0,1]$  as well as no $\chi \in \left[0,1\right]$ such that

\begin{equation}
    \chi \mu^{\prime}(\gamma) + (1-\chi) \mu^{\prime \prime} (\gamma) < \mu_{\text{H}} = \mu + 2 \frac{|\delta|}{D} = \frac{Q_{11} + Q_{22} + 2|\delta|}{\Delta},
    \label{eq:thm_2_conditions}
\end{equation}
\noindent where $\Delta {=} \text{det}[\textbf{Q}]{=}Q_{11}Q_{22}{-}Q_{12}^2 $. Notice that in the proof of the previous theorem, we implicitly assumed $\chi {=}\frac{1}{2}$ as we added up the two bounds in Eq.~\eqref{eq:summed_up} with equal weights.  Now for QFIM close to diagonal, we can do the following perturbation expansions
\begin{align}
    \frac{1}{Q_{11}}{=} \frac{Q_{22}}{\Delta}\left(1+\frac{Q_{12}^2}{\Delta} \right)^{-1} \approx \frac{Q_{22}}{\Delta}\left(1-\frac{Q_{12}^2}{\Delta}\right) \\
    \frac{1}{Q_{22}}{=} \frac{Q_{11}}{\Delta}\left(1+\frac{Q_{12}^2}{\Delta} \right)^{-1} \approx \frac{Q_{11}}{\Delta}\left(1-\frac{Q_{12}^2}{\Delta}\right)
\end{align}

Plugging in these terms, after some algebra the condition in inequality ~\eqref{eq:thm_2_conditions} above reads 
\begin{align}
    \chi(1-2\gamma)(Q_{11}-Q_{22}) -\left[\gamma^2 Q_{22} + (1-\gamma)^2 Q_{11} \right] \nonumber \\
    + 2\gamma(1-\gamma)|\delta| + \frac{\gamma Q_{12}^2 \left[ \chi Q_{11} + (1-\chi)Q_{22}\right]}{\Delta} < 0
\end{align}

Now, let us search for the $(\gamma,\chi)$ tuple that may violate this inequality above. If without loss of generality we assume that $Q_{11} > Q_{22}$, then it is clear that the LHS is maximized, i.e., we give ourselves the best chance of violating this inequality, when $\chi{=}1$. In this case, the inequality above reads $f(\gamma) <0 $, where

\begin{align}
    f(\gamma) = (1-2\gamma)(Q_{11}-Q_{22}) -\left[\gamma^2 Q_{22} + (1-\gamma)^2 Q_{11} \right] \nonumber \\
    + 2\gamma(1-\gamma)|\delta| + \frac{\gamma Q_{12}^2  Q_{11}}{\Delta} 
\end{align}
\noindent By maximizing with respect to $\gamma$ through putting $\partial f/\partial\gamma =0$ (one can check $\partial^2f{/}\partial^2 \gamma {<} 0$), the maximum value of $f(\gamma) = f_{\max}$ is given by 

\begin{equation}f_{\max}=\frac{Q_{11}^2 Q_{12}^4+4 Q_{11} \Delta \left[Q_{11} Q_{22}^2-Q_{12}^2 (\delta +2 Q_{22})\right]+4 \Delta ^2 |\delta|^2}{4 \Delta^2 (2 |\delta| +Q_{11}+Q_{22})} 
\label{eq:necessary}
\end{equation}

\noindent Now one can write $Q_{12} = \epsilon \sqrt{Q_{11}Q_{22}}$, and as this off-diagonal element is relatively small, one can expand $f_{\max}$ as a power series in $\epsilon$ in the following manner

\begin{equation}
    f_{\max} = \frac{|\delta|^2 - Q_{11}Q_{22}}{Q_{11}+Q_{22}+2|\delta|} + \epsilon^2 \frac{Q_{11} (Q_{22}+|\delta|)}{Q_{11}+Q_{22}+2|\delta|} + \mathcal{O}(\epsilon^4)
\end{equation}

However, one can prove using the Cauchy-Schwarz inequality on their respective definitions that $Q_{11}Q_{22} > |\delta|^2$. Therefore, upto leading order, this inequality in Eq.~\eqref{eq:thm_2_conditions} is always satisfied.  Therefore, for QFIM that are very close to diagonal, JE always remains a better strategy compared to stepwise strategy, even if we consider the saturable HCRB for joint estimation. \qed \\

However, we note that this is \emph{only just}, since the presence of the Holevo term lends a significant boost to the possibility of SE outperforming JE if we deviate from this almost diagonal QFIM paradigm. Indeed, keeping terms up to one more subleading order, we already have the following necessary condition of SE outperforming JE HCRB. 

\begin{equation}
    \frac{Q_{12}^2}{Q_{11}Q_{22}} \geq \frac{ 1 - \frac{|\delta|^2}{Q_{11} Q_{22}}}{1 + \frac{|\delta|}{Q_{22}}}
\end{equation}

\begin{figure}
    \centering
    \includegraphics[width=0.5\linewidth]{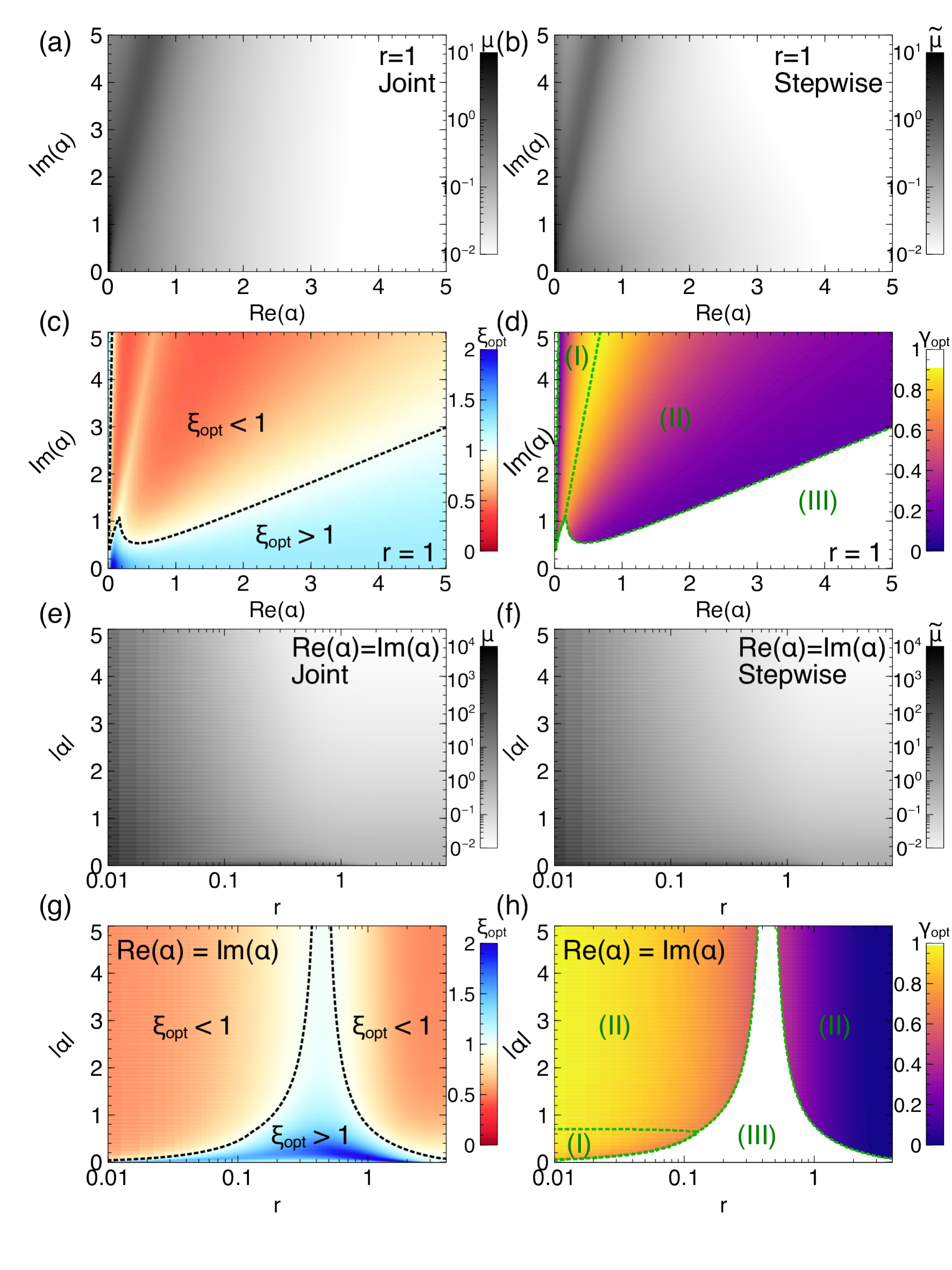}
    \caption{Coherent-state probe estimating $(\phi, r)$. (a) JE bound $\mu$ with initial displacement $(\text{Re}(\alpha), \text{Im}(\alpha))$, where $r{=}1$. (b) same for optimal SE bound $\tilde{\mu}$.(c) Ratio $ \xi_{\rm{opt}}=\tilde{\mu}/\mu$, black dotted line separates regions where JE (blue) or SE (red) furnishes a better bound $(\tilde{\mu}/\mu {=}\xi_{\text{opt}} {\gtrless} 1)$. (d) Optimal strategy $\gamma_{\text{opt}}$, green dotted lines separate regions (I) SE estimating $\phi$ first, (II) SE estimating $r$ first, (III) JE, furnishes a lower bound. (e)-(h) reproduces same figures as (a)-(d) with parameter values $r$ and initial parameter $|\alpha|$ along $\text{Re}(\alpha){=}\text{Im}(\alpha)$ direction. }
    \label{fig:cv}
\end{figure}

\subsection{Gaussian probes} 

Let us now illustrate the performance of SE in a continuous-variable multiparameter estimation problem. Firstly, let us note one common feature of SE strategies When the QFIM is diagonal - namely, they result in a worse bound than JE. This result follows trivially by choosing $\textbf{Q} {=} \text{diag}(Q_{11},Q_{22})$, in which case for both strategies $\mu' {=}\mu'' {=} \sum_{i} 1/m_{i}Q_{ii} {\geq} 1/\sum_{i}m_{i}Q_{ii}{=}\mu$. Thus, in some situations in Gaussian quantum metrology where QFIM is nearly diagonal \cite{nichols2018multiparameter}, SE is unlikely to achieve any operational advantage beyond the usual symmetric logarithmic derivative version of JE. Specifically, multiparameter estimation problems with a displaced squeezed state probe fall into this class, where one seeks to simultaneously estimate an unknown displacement parameter $\textbf{d}$ and an unknown real squeezing parameter $r$. However, let us consider an initially coherent state probe $|\psi(\alpha)\rangle$ undergoing a phase rotation $\hat{R}(\phi)$ following a squeezer $\hat{S}(r)$. The final state of the probe is $|\psi_{\text{probe}}\rangle {=} \hat{R}(\phi)\hat{S}(r) |\psi(\alpha)\rangle$, from which both rotation angle $\phi$ and squeezing parameter $r$ are to be sensed. The QFIM for this model reads ~\cite{bakmou2020multiparameter} 
\begin{equation}
    \textbf{Q} = \begin{bmatrix}
        8 |\alpha|^2 + 2 \tanh^2(4r) & -16 \text{Re}(\alpha)\text{Im}(\alpha)\cosh(2r) \\
        -16 \text{Re}(\alpha)\text{Im}(\alpha)\cosh(2r)  & 8e^{4r}\text{Re}(\alpha)^2 + 8 e^{-4r} \text{Im}(\alpha)^2 
    \end{bmatrix}
    \label{eq:qfim_gaussian_bakmou}
\end{equation}
\noindent This expression for QFIM is independent of phase rotation angle $\phi$ and solely depends on the squeezing parameter $r$. As illustrated in Fig.~\ref{fig:cv}(a)-(d), SE offers considerable more precision especially where the initial coherent state is displaced more along the one quadrature direction than another. Fig.~\ref{fig:cv}(e)-(h) demonstrates that choosing an initial displacement along heterodyne directions results in SE becoming more advantageous to sense both low and high magnitudes of squeezing parameter $r$, with JE furnishing a better bound only in the intermediate regime.\\




\subsection{Bayesian Scheme for Stepwise Quantum Estimation}

Here, we detail the Bayesian scheme for stepwise quantum estimation. For concreteness we assume $\lambda_1$ is measured first, then the Bayesian estimate of $\lambda_1$ is fed back to measure $\lambda_2$. Let us assume the priors are given by $p_1(\lambda_1)$ and $p_2(\lambda_2)$ respectively. Now, we proceed sequentially in the following manner.

\begin{enumerate}
    \item \textbf{Estimating $\lambda_1$ --} Let us assume the measurement result is $x_1$. For example, for qubits, this can be a numerically simulated by a binary string randomly picked via binomial probabilities given by the true parameter. The posterior of the first parameter is then given by the Bayes rule as 
    \begin{eqnarray}
        p_1(\lambda_1|x_1) \propto \int p_1(\lambda_1|x_1,\lambda_2) p_2(\lambda_2) d\lambda_2 \nonumber\\
        = \int p_1(x_1|\lambda_1,\lambda_2) p_1(\lambda_1|\lambda_2)p_2(\lambda_2) d\lambda_2 \nonumber \\
        = \left[\int p_1(x_1|\lambda_1,\lambda_2) p_2(\lambda_2) d\lambda_2 \right] p_{1}(\lambda_1)
    \end{eqnarray}
    \noindent Note that the likelihood function involves averaging over the prior $p_2$ of the parameter $\lambda_2$. After normalizing this posterior, we can now obtain the Bayesian estimate $\lambda_1^{est}$ of $\lambda_1$ as the weighted average over posterior, i.e., $\lambda_1^{est}{=}\sum\lambda_1p_{1}(\lambda_1|x_1)$, and the variance $\Delta^2_{\lambda_1} = \left[\sum\lambda_1^2 p_{1}(\lambda_1|x_1)\right]- (\lambda_1^{est})^2$. \\
   \item \textbf{Estimating $\lambda_2$ --} Armed with the Bayesian estimate of the first parameter, now we can update the prior $p_2$ of the second parameter $\lambda_2$ after another round of simulated measurements $x_2$ given by another binary string randomly picked via binomial probabilities given by the true parameter values. Typically this measurement is performed in a different basis than the first step. Now, the posterior of the probability $p_2$ for the second parameter $\lambda_2$ is given by the Bayes rule as   \begin{eqnarray}
        p_2(\lambda_2|x_2) \propto \left[  p_2(x_2|\lambda_1^{est},\lambda_2)\right] p_{2}(\lambda_2)
        \end{eqnarray}
        \noindent  After normalizing this posterior, we can now obtain the Bayesian estimate $\lambda_2^{est}$ of $\lambda_2$ as the weighted average over posterior, i.e., $\lambda_2^{est}{=}\sum\lambda_2p_{2}(\lambda_2|x_2)$, and the variance $\Delta^2_{\lambda_2} = \left[\sum\lambda_2^2 p_{2}(\lambda_2|x_2)\right]- (\lambda_2^{est})^2$. 
\end{enumerate}

\noindent Notice that within each round of simulated run of measurements, the length of outcome strings can be different for two parameters, referring to $m_1$ and $m_2$ proportional allocations in the main paper. As more and more rounds are performed, the Bayesian estimates converge to their true values. At this point, we remind the reader that this is \emph{not} adaptive, since measurement settings and these proportions $(m_1,m_2)$ for both parameters are kept fixed at each round. \\

\end{document}